\documentclass{ws-procs961x669} 

\begin{document}
\title{Wormhole interaction in $2$d Hor\v ava-Lifshitz quantum gravity}

\author{Jan Ambj\o rn$^a$}

\address{The Niels Bohr Institute, Copenhagen University,\\
Blegdamsvej 17, DK-2100 Copenhagen, Denmark\\
and\\
IMPP, Radboud University,\\
Heyendaalseweg 135, 6525 AJ, Nijmegen, The Netherlands\\
$^a$E-mail: ambjorn@nbi.dk}

\author{Yuki Hiraga$^b$, Yoshiyasu Ito$^c$ and Yuki Sato$^d$}

\address{Department of Physics, Nagoya University,\\
Chikusaku, Nagoya 464-8602, Japan\\
$^b$E-mail: hiraga@eken.phys.nagoya-u.ac.jp\\
$^c$E-mail: ito@eken.phys.nagoya-u.ac.jp\\
and\\
Institute for Advanced Research, Nagoya University,\\
Chikusaku, Nagoya 464-8602, Japan\\
$^d$E-mail: ysato@th.phys.nagoya-u.ac.jp
}

\begin{abstract}
A lattice regularization for the $2$d projectable Hor\v ava-Lifshitz (HL) quantum gravity is known to be the $2$d causal dynamical triangulations (CDT), 
and the $2$d CDT can be generalized so as to include all possible genus contributions non-perturbatively. 
We show that in the context of HL gravity, 
effects coming from such a non-perturbative sum over topologies can be successfully taken into account, 
if we quantize the $2$d projectable HL gravity with a simple bi-local wormhole interaction. 
This conference paper is based on the article, Phys. Lett. B \textbf{816} (2021), 136205\cite{Ambjorn:2021wou}.      
\end{abstract}


\bodymatter

\section{Introduction}\label{sec:introduction}

Two-dimensional models of quantum gravity are useful for understanding some aspects of non-perturbative physics, 
since they can be examined beyond the framework of perturbation theory by analytic computations in many cases. 
One of such models is the two-dimensional causal dynamical triangulations ($2$d CDT)\cite{Ambjorn:1998xu}, 
which is a lattice toy model of quantum gravity with a global time foliation. 
Global hyperbolicity imposed at the quantum level do not allow for any topology change in the CDT model, 
and the continuum limit can be described by  quantum mechanics of a $1$d universe\cite{Ambjorn:1998xu}. 

Yet another toy model for quantum gravity that has a global time foliation is  $2$d Hor\v ava-Lifshitz (HL) quantum gravity 
that was introduced first in higher dimensions to resolve the issue of perturbative renormalizability by breaking the diffeomorphisms down to the foliation-preserving diffeomorphisms\cite{horava}, 
and so the model has a preferred foliation structure. 
The projectable version of the $2$d HL quantum gravity was discussed in the articles\cite{Ambjorn:2013joa, Li:2014bla}, and in particular it was shown that 
the canonical quantization of the model yields exactly the same quantum Hamiltonian as the one obtained by the continuum limit of $2$d CDT\cite{Ambjorn:2013joa}. 
Therefore, one can interpret $2$d CDT as a lattice regularization for the $2$d projectable HL quantum gravity.    

Although the creation of baby universes and wormholes (handles) is not allowed to occur in the original setup of CDT, 
one can generalize the $2$d model to include such configurations in a manner consistent with the scaling limit\cite{Ambjorn:2007jm}, 
and such a generalized model can be fully described by a string field theory for CDT\cite{Ambjorn:2008ta}\footnote{The time-independent amplitudes can be also computed by the matrix model for CDT\cite{Ambjorn:2008jf} and by the new scaling limit of the Hermitian one-matrix model\cite{Ambjorn:2008gk}.}. 
Here the word ``string'' refers to the $1$d spatial universe. 
Based on the string field theory, one can take into account all genera (handles) as well as all baby universes of the 
two-dimensional spacetime. 
In addition, as long as one only asks for the amplitude between the states describing  connected 
spatial universes separated a proper time $t$,  this sum over topologies and baby universes can effectively be described
by a one-body Hamiltonian\cite{Ambjorn:2009fm,Ambjorn:2009wi}.         

Having the equivalence between the continuum limit of $2$d CDT and the $2$d projectable HL quantum gravity, 
we discussed in the article\cite{Ambjorn:2021wou} what kind of effective interaction should be added to the classical Lagrangian of the $2$d projectable HL gravity when canonically quantized,
in order to obtain the above mentioned one-body quantum Hamiltonian that includes all wormholes and baby universes obtained in $2$d CDT.  
The answer is that it is enough to include a 
simple bi-local and spatial wormhole interaction that is compatible with the foliation-preserving diffeomorphisms.

This article is organized as follows. 
In Section \ref{sec:cdtandhl}, we give a brief introduction to  $2$d CDT and  $2$d HL quantum gravity, 
and explain the relation between the two. 
In  Section \ref{sec:sumover} the string field theory for CDT is described and we introduce the one-body quantum Hamiltonian that includes all possible topologies. 
In Section \ref{sec:wormholeinhl}, we introduce a simple wormhole interaction to the $2$d projectable HL gravity and show that one can precisely recover the one-body Hamiltonian when canonically quantizing the model. 
Section \ref{sec:discussions} is devoted to discussions.

\section{CDT and Hor\v ava-Lifshitz gravity in two dimensions}\label{sec:cdtandhl}

In this Section, we review causal dynamical triangulations (CDT) and Hor\v ava-Lifshitz gravity (HL) in two dimension, 
and explain the relation between the two. 

\subsection{$2$d CDT}\label{sec:2dcdt}  
The starting point is a globally hyperbolic manifold equipped with a global time foliation:
\begin{equation}
\mathcal{M} = \bigcup_{t\in \mathbb{R}} \Sigma_t\ , 
\label{eq:foliation}
\end{equation}
where each leaf $\Sigma_t$ is a Cauchy ``surface''. 
 $2$d CDT is a model which quantizes $2$d geometries with such a proper-time foliation. 
The geometries in the path integral are regularized by piecewise linear geometries constructed by gluing together 
special kinds of triangles \cite{Ambjorn:1998xu}. 
Each triangle consists of one space-like edge and two time-like edges such that 
a square of the space-like edge is positive, $a^2_s = \varepsilon^2$, while it is negative for the time-like edge, $a^2_t = - \alpha \varepsilon^2$ with positive $\alpha$.

If we prohibit a creation of baby universes, a reasonable choice of the action is the cosmological constant term that is regularized as follows:
\begin{align}
&S[g_{ij}]=-\lambda_0 \int d^2x\ \sqrt{-\text{det}(g_{ij})}\ , \notag \\
&\ \ \  \rightarrow \ \ \ 
S_R [T;\alpha] = -\frac{\mu}{\varepsilon^2} \times 
\left( 
\frac{\sqrt{4\alpha + 1}}{4} \varepsilon^2
\times n(T)
\right)\ ,
\label{eq:regge}
\end{align}
where the bare cosmological constant $\lambda_0$ is replaced by the dimensionless cosmological constant $\mu$, 
and the quantity in the parentheses is the discretized area of the triangulation $T$, i.e. 
the  area of each triangle times the number of triangles $n(T)$. 
The lattice action $S_R$ in eq.\ (\ref{eq:regge}) is called the Regge action \cite{Regge:1961px}.      

For computational convenience we implement a rotation to the Euclidean signature which can be done by replacing $\alpha$ with $-\alpha$:
\begin{equation}
iS_R [T;\alpha] \to -S^{(e)}_R [-\alpha;T] = - \mu \frac{\sqrt{4\alpha - 1}}{4} n(T) =: - \mu\, n(T)\ ,
\label{eq:euclideanregge}
\end{equation}
where we have absorbed a numerical factor into the dimensionless cosmological constant. 
Note that this map is a bijection between individual Lorentzian and Euclidean geometries.  

In CDT, the integration over diffeomorphism equivalent classes of metric $g$ keeping both initial and final geometries fixed 
can be regularized by the sum over ``all'' triangulations. 
Therefore, the $2$d Euclidean path-integral regularized by CDT is 
\begin{equation}
G^{(0)}_{\text{lattice}} (l_1,l_2;\tau) 
=\sum_{T \in \mathcal{T}(l_1,l_2;\tau)} \frac{1}{C_T} e^{-\mu n(T)}\ , 
\label{eq:glattice}
\end{equation}
where $C_T$ is the order of automorphism group of $T$, 
and $\mathcal{T}(l_1,l_2;\tau)$ is a set of triangulations such that the initial and final boundaries whose lengths are kept fixed to $l_1$ and $l_2$ 
are separated by $\tau$  Euclidean time steps. 

One can compute the amplitude (\ref{eq:glattice}) analytically through the use of the generating function for the numbers $G^{(0)}_{\text{lattice}} (l_1,l_2;\tau)$ \cite{Ambjorn:1998xu}. 
As in the case of lattice QCD, tuning the UV relevant coupling constant $\mu/\varepsilon^2$ to its critical value $\mu_c/\varepsilon^2$ and taking $\varepsilon \to 0$ in a correlated manner, 
one can transmute the dimension of the lattice spacing to the renormalized coupling constant:
\begin{equation}
\lambda := \lim_{\substack{\varepsilon \to 0 \\ \mu \to \mu_c}}  \frac{\mu - \mu_c}{\varepsilon^2}\ ,
\label{eq:lambda}
\end{equation}    
where $\lambda$ is the renormalized cosmological constant. 
Introducing the renormalized quantities, boundary lengths and a proper time such that
\begin{equation}
\ell_1 := \varepsilon l_1\ , \ \ \ \ell_2 := \varepsilon l_2\ , \ \ \ t := \varepsilon \tau\ ,
\label{eq:renormalized}
\end{equation}   
one can obtain the renormalized amplitude $G^{(0)}(\ell_1,\ell_2;t)$ that is known to satisfy 
the differential equation \cite{Ambjorn:1998xu, Ambjorn:2013joa}:
\begin{equation}
- \frac{\partial}{\partial t} G^{(0)}(\ell_1,\ell_2;t) 
= H^{(0)}_a (\ell_1) G^{(0)}(\ell_1,\ell_2;t)\ , 
\label{eq:differentialeq}
\end{equation}
where $H^{(0)}_a$ is the quantum Hamiltonian defined as
\begin{equation}
H^{(0)}_{0} (\ell)
= - \frac{\partial}{\partial \ell} \ell \frac{\partial}{\partial \ell} + \lambda \ell\ , \ \ \ 
H^{(0)}_{-1} (\ell)
= - \ell \frac{\partial^2}{\partial \ell^2}  + \lambda \ell\ , \ \ \ 
H^{(0)}_{+1} (\ell)
= - \frac{\partial^2}{\partial \ell^2} \ell + \lambda \ell\ .
\label{eq:hamiltonians}
\end{equation}
Here the label $a$ in eq.\ (\ref{eq:differentialeq}) specifies the ordering of the Hamiltonian. 
If we define a quantum state of the one-dimensional universe whose length is $\ell$ as $|\ell \rangle$, 
the amplitude $G^{(0)}(\ell_1,\ell_2;t)$ can be rewritten as 
\begin{equation}
G^{(0)}(\ell_1,\ell_2;t) = \langle \ell_2 | e^{-t H^{(0)}_a (\ell_1)} | \ell_1 \rangle\ .
\label{eq:operatorh}
\end{equation}
In fact, each ordering specifies the geometry of the quantum state $|\ell \rangle$, i.e. 
if $a=0,-1,+1$, then the geometry of the one-dimensional universe is open, closed with a mark, and closed, respectively \cite{Ambjorn:1998xu, Ambjorn:2013joa}.  
Marking a point on a closed universe is analogous to introducing a coordinate. 
The Hamiltonian is hermitian with respect to the following inner product:
\begin{equation}
\langle \phi | H^{(0)}_a | \psi \rangle 
= \int^{\infty}_0 \phi^* (\ell) H^{(0)}_a (\ell) \psi (\ell)\ d\mu_a(\ell)\ , \ \ \ 
\text{with}\ \ \ 
d\mu_a(\ell) = \ell^ad\ell\ . 
\label{eq:innerproduct}
\end{equation} 

As a result, the physics of $2$d CDT can be described by the quantum mechanics of a one-dimensional universe.

\subsection{$2$d projectable HL gravity}\label{sec:2dhl}  

The starting point is the same as that of CDT, i.e. a globally hyperbolic manifold equipped with a global time foliation (\ref{eq:foliation}). 
A natural parametrization for the metric on such a geometry is given by the Arnowitt-Deser-Misner metric:
\begin{equation}
g = -N^2ds^2+ h_{11} (dx+N^1ds)(dx+N^1ds)\ ,
\label{eq:adm}
\end{equation}
where $h_{11}$ is the spatial metric on the leaf; $N$ and $N^1$ called lapse and shift functions, 
quantify the normal and tangential directions of the proper time to the leaf, respectively.  

The $2$d HL gravity is introduced as a theory that keeps the foliation structure \cite{horava}, 
or in other words, it is invariant under the foliation preserving diffeomorphisms (FPD):
\begin{equation}
s \to s + \xi^0(s)\ , \ \ \ x\to x+ \xi^1 (s,x)\ . 
\label{eq:fpd}
\end{equation}
Under the FPD, the fields transform as
\begin{align}
\delta_{\xi} h_{11} 
&= \xi^0 \partial_0 h_{11} + \xi^1 \partial_1 h_{11} + 2 h_{11}\partial_1 \xi^1
\ , 
\label{eq:fpd1}
\\
\delta_{\xi} N_1 
&=
\xi^{\mu} \partial_{\mu} N_1 +N_1 \partial_{\mu} \xi^{\mu} +h_{11} \partial_0 \xi^1\ , 
\label{eq:fpd2}
\\
\delta_{\xi} N 
&= \xi^{\mu} \partial_{\mu} N + N \partial_0 \xi^0\ ,
\label{eq:fpd3}
\end{align}
where $N_1 = h_{11}N^1$.

Note that if the lapse function $N$ is a function of time, $N=N(t)$, it stays as a function of time under the FPD. 
The $2$d projectable HL gravity satisfies this condition on $N$, and it is defined by the following action: 
\begin{equation}
I =
\frac{1}{\kappa}
 \int dsdx\ N\sqrt{h}
\left(
(1-\eta)K^2 -2\tilde\lambda 
\right)\ ,
\label{eq:actionI}
\end{equation}
where $\eta$, $\tilde\lambda$ and $\kappa$ are a dimensionless parameter, 
the cosmological constant and the dimensionless gravitational coupling constant, respectively; 
$h$ is the determinant of the metric $h_{11}$, i.e. $h=h_{11}$;  
$K$ is the trace of the extrinsic curvature $K_{11}$ defined as 
\begin{equation}
K_{11} = \frac{1}{2N} 
\left( 
\partial_0 h_{11} 
-2\nabla_1 N_1
\right)\ , \ \ \ \text{with}\ \ \ \nabla_1 N_1 := \partial_1 N_1 - \Gamma^1_{11} N_1\ . 
\label{eq:extrinsiccurv}
\end{equation}
Here $\Gamma^1_{11}$ is the spatial Christoffel symbol:  
\begin{equation}
\Gamma^1_{11} = \frac{1}{2} h^{11}\partial_1 h_{11}\ .  
\label{eq:Christoffel }
\end{equation}
In principle, one can add higher spatial derivative terms to the action (\ref{eq:actionI}), but they are not needed since 
2d gravity is renormalizable without such terms and we will omit such terms.

The quantization of $2$d projectable HL gravity was discussed in \cite{Ambjorn:2013joa, Li:2014bla}, 
and in particular, it was shown that the quantum Hamiltonian coincides with the continuum Hamiltonian of $2$d CDT when the following identification of the parameters is made\cite{Ambjorn:2013joa}:
\begin{equation}
\lambda = \frac{\tilde \lambda}{2(1-\eta)}\ , \ \ \ \eta < 1\ , \ \ \ \tilde\lambda > 0\ , \ \ \ \kappa = 4(1-\eta)\ .
\label{eq:lambdacdt}
\end{equation}  
where $\lambda$ is the renormalized cosmological constant in $2$d CDT (\ref{eq:lambda})\footnote{We have set unimportant dimensionless gravitational constant as $\kappa = 4(1-\eta)$.}. 

Let us briefly explain how to recover the quantum Hamiltonian (\ref{eq:hamiltonians}) from the quantization of $2$d HL gravity. 
Introducing the conjugate momentum of $\sqrt{h}$ as $\pi$, we have in  the canonical formalism the Poisson bracket  
\begin{equation}
\left\{ \sqrt{h(s,x)}, \pi (s,x') \right\} = \delta (x-x')\, ,
\label{eq:PB}
\end{equation}
and corresponding to the Lagrangian  (\ref{eq:actionI}) we have the Hamiltonian
\begin{equation}
H = \int dx\ \left[
N_1 \left(
- \frac{\partial_1 \pi}{\sqrt{h}} 
\right)
+N \left(
\frac{\kappa}{4(1-\eta)} \pi^2 \sqrt{h} 
+ \frac{2}{\kappa}\tilde \lambda \sqrt{h} 
\right)
\right]\, .
\end{equation}

If we solve the momentum constraint at the classical level, i.e.
\begin{equation}
- \frac{\partial_1 \pi}{\sqrt{h}} = 0\ , \ \ \ \Rightarrow \ \ \ \pi = \pi (s)\ ,
\label{eq:momentumconst}
\end{equation}
the system reduces to a one-dimensional model with the Hamiltonian
\begin{equation}
H= N(s) \left( \frac{\kappa}{4(1-\eta)} \pi^2(s) \ell (s) + \frac{2}{\kappa} \tilde \lambda \ell (s)  \right)\ , \ \ \ 
\text{with}\ \ \ \ell (s) := \int dx\ \sqrt{h(s,x)}\ .
\label{eq:1dhamiltonian}
\end{equation}
Hereafter choosing the correct sign for the kinetic term, i.e. $\eta<1$, 
we use the parametrization (\ref{eq:lambdacdt}) with positive $\lambda$ in order to discuss the relation to $2$d CDT.

The classical $1$d system with the Hamiltonian (\ref{eq:1dhamiltonian}) can be alternatively described by the following action: 
\begin{equation}
S
= 
\int^{1}_{0} ds
\left(
\frac{\dot \ell^2}{4 N\ell} 
- \lambda N\ell  
\right)\ ,
\label{eq:actionL}
\end{equation}
where $\dot \ell := d\ell/ds$.  
This system is invariant under the time reparametrization, $s \to s + \xi^0(t)$, 
which is ensured by the lapse function. 
In fact, the proper time, 
\begin{equation}
t := \int^1_0 ds\ N(s)\ , 
\label{eq:propertime}
\end{equation}
and the length, $\ell=\ell(t)$, are invariant under the time reparametrization, 
and so it makes sense to discuss the probability amplitude for a $1$d universe to propagate in the proper time $t(>0)$, 
starting from the state with the length $\ell_1$ and ending up in the one with  length $\ell_2$ \cite{Ambjorn:2013joa}.  
Such an amplitude can be computed based on the path-integral, and we evaluate it by a rotation to the Euclidean signature for convenience.  
In our foliated spacetime, for $\eta < 1$, we can implement this procedure by a formal rotation, $s\to is$, which yields the amplitude:
\begin{equation}
G^{(0)}(\ell_2,\ell_1;t) 
= \int \frac{\mathcal{D}N(s)}{\text{Diff}[0,1]} \int^{\ell (1) = \ell_2}_{\ell (0)=\ell_1} \mathcal{D}\ell(s) e^{-S_E [N(s),\ell(s)]}\ ,
\label{eq:cylinder}
\end{equation}
where $\text{Diff}[0,1]$ is the volume of the time reparametrization; 
$S_E$ is the Euclidean action given by
\begin{equation}
S_E
= 
\int^{1}_{0} ds 
\left(
\frac{\dot \ell^2}{4N\ell} 
+ \lambda N\ell 
\right)\ ,
\label{eq:actionE}
\end{equation}
where $\dot \ell := d\ell/ds$.

We set $N=1$ as a gauge choice. 
One can show that the corresponding Faddeev-Popov determinant only gives an overall constant, 
which we will omit in the following. 
The amplitude (\ref{eq:cylinder}) then becomes
\begin{equation}
G^{(0)}(\ell_2,\ell_1;t) 
= \int^{\ell (t) = \ell_2}_{\ell (0)=\ell_1} \mathcal{D}\ell (s) 
\exp \left[ 
- \int^{t}_{0} ds 
\left(
\frac{\dot \ell^2}{4\ell} 
+ \lambda \ell
\right)
\right]
\ ,
\label{eq:cylinder2}
\end{equation}
which can be expressed in terms of the quantum Hamiltonian $H$ that is unknown at the moment:
\begin{equation}
G^{(0)}(\ell_2,\ell_1;t) 
=
\langle \ell_2 | e^{-t H} | \ell_1\rangle\ ,
\label{eq:cylinder3}
\end{equation}
where $| \ell \rangle$ is a quantum state of the $1$d universe with the length $\ell$. 
By standard methods (see e.g. the article\cite{Ambjorn:2013joa}), one can determine the quantum Hamiltonian from eq.\ (\ref{eq:cylinder2}) and eq.\ (\ref{eq:cylinder3}). 
One obtains  precisely the  $2$d CDT quantum Hamiltonians given by eq.\ (\ref{eq:hamiltonians}) if  
the integral measures are chosen as
\begin{equation}
\mathcal{D}\ell (s) = \prod^{s=t}_{s=0} \ell^a(s)d\ell (s)\ ,
\label{eq:measure2}
\end{equation}
where $a=0,\pm 1$. 
The measures (\ref{eq:measure2}) are consistent with eq.\ (\ref{eq:innerproduct}) introduced in $2$d CDT.

Therefore, we can conclude that $2$d CDT is a lattice regularization for the $2$d projectable HL quantum gravity.

\section{Sum over all genera in $2$d CDT}\label{sec:sumover}

One can generalize the $2$d CDT model so as to include spatial topology changes (splitting and joining interactions of $1$d universe) in keeping with the foliation structure,  
and this is described by promoting the $1$d quantum mechanics to a field theory of $1$d universes, 
which is called the string field theory for CDT \cite{Ambjorn:2008ta}. Here the word ``string'' means a $1$d closed spatial 
universe. 

We introduce an operator that creates a marked closed string with the length $\ell$, $\Psi^{\dagger} (\ell)$, 
and an operator that annihilates a length $\ell$ closed string without a mark, $\Psi (\ell)$. 
They satisfy the commutation relation:
\begin{equation}
[ \Psi (\ell), \Psi^{\dagger} (\ell') ] 
= \delta (\ell-\ell'), \quad [ \Psi (\ell), \Psi(\ell') ] = [ \Psi^\dagger (\ell), \Psi^{\dagger} (\ell') ] = 0.
\label{eq:commutator}
\end{equation} 
The vacuum $|\text{vac} \rangle$ is defined by $0=\Psi (\ell) |\text{vac} \rangle = \langle \text{vac} | \Psi^{\dagger} (\ell)$.  

The CDT amplitude (\ref{eq:operatorh}) can be written in terms of the string field Hamiltonian 
obtained by sandwiching the one-body Hamiltonian:
\begin{equation}
G^{(0)} (\ell_1, \ell_2 ;t)
= \langle \text{vac} | \Psi (\ell_2)\ e^{-t \hat{H}^{(0)}} \Psi^{\dagger} (\ell_1) | \text{vac} \rangle\ , 
\label{eq:secondg}
\end{equation}
where
\begin{equation}
\hat{H}^{(0)} 
= \int^{\infty}_{0} \frac{d\ell}{\ell} \Psi^{\dagger} (\ell) H^{(0)}_{-1} \Psi (\ell)\ .
\label{eq:secondhfree} 
\end{equation}

In order to incorporate topology change, one has to include suitable interactions, 
and the full string field Hamiltonian is given by\cite{Ambjorn:2008ta}:
\begin{align}
\hat H 
&= \hat H^{(0)} 
- \int^{\infty}_{0} d\ell \ \delta (\ell) \Psi (\ell)
-g_s \int^{\infty}_0 d\ell_1 \int^{\infty}_0 d\ell_2\ \Psi^{\dagger} (\ell_1) \Psi^{\dagger} (\ell_2) (\ell_1+\ell_2) \Psi (\ell_1 + \ell_2) \notag \\
& \ \ \ 
-\alpha g_s \int^{\infty}_{0} d\ell_1 \int^{\infty}_0 d\ell_2\ \Psi^{\dagger} (\ell_1 + \ell_2) \ell_1 \Psi(\ell_1) \ell_2 \Psi (\ell_2)\ , 
\label{eq:fullh}
\end{align}
where the second, third and fourth terms mean 
a string vanishing into the vacuum, the splitting interaction with the string coupling $g_s$, 
and the joining interaction with the coupling $\alpha g_s$, respectively;  
$\alpha$ is a constant introduced for counting handles. 
In general one can compute the following amplitude:
\begin{equation}
A (\ell_1, \cdots, \ell_m ; \ell'_1, \cdots, \ell'_n;t) 
= \langle \text{vac} | \Psi (\ell'_1) \cdots \Psi (\ell'_n)
e^{-t \hat H} 
\Psi^{\dagger} (\ell_1) \cdots \Psi^{\dagger} (\ell_m) 
| \text{vac} \rangle\ .
\label{eq:generalamplitude}
\end{equation}

From now on, we focus on the full propagator $G (\ell_1, \ell_2 ;t) (:= A (\ell_1; \ell_2 ;t))$ that includes the sum over all genera (handles) and baby universes, 
and so we simply set $\alpha =1$. 
In this case, somewhat miraculously,  
the full propagator in the multi-body system can be  described by an effective one-body system\cite{Ambjorn:2009fm,Ambjorn:2009wi}:
\begin{equation}
G (\ell_1, \ell_2 ;t) 
= \langle \ell_2 | e^{-t H_{-1}} | \ell_1 \rangle\ , 
\ \ \ \text{with}\ \ \ 
H_{-1} = -\ell \frac{\partial^2}{\partial \ell^2} + \lambda \ell - g_s \ell^2\ .
\label{eq:onebodyfullh}
\end{equation}
All the contributions coming from the sum over all genera and baby universes 
can be effectively described by the last term $-g_s\ell^2$. 
Although the Hamiltonian (\ref{eq:onebodyfullh}) is not bounded from below, 
it belongs to a class of Hamiltonians which are called ``classical incomplete'', 
where the Hamiltonians have  discrete energy spectra and square integrable eigenfunctions. 

The effective one-body system given by eq.\ (\ref{eq:onebodyfullh}) can be also described by the path-integral\cite{Ambjorn:2021wou, Ambjorn:2021wdm}: 
\begin{equation}
G(\ell_1,\ell_2;t)
= \int^{\ell (t) = \ell_2}_{\ell (0)=\ell_1} \mathcal{D}\ell (s)\ 
\exp 
\left[
- \int^t_0 ds\ 
\left(
\frac{\dot \ell^2(s)}{4\ell(s)}
+ \lambda \ell(s)
-g_s \ell^2 (s)
\right)
\right]\ .
\label{eq:pathintegralfullg}
\end{equation}    
If we choose the integral measure such that 
\begin{equation}
\mathcal{D}\ell (s) = \prod^{s=t}_{s=0} \ell^a(s)d\ell (s)\ , 
\label{eq:measure3}
\end{equation}
where $a=0,\pm1$, 
all possible orderings of the full one-body Hamiltonian (\ref{eq:onebodyfullh}) can be realized.  
As explained in the article\cite{Ambjorn:2021wdm}, in order for the functional integral to be well defined, the boundary conditions on $\ell (s)$ at infinity have to be chosen such that 
the kinetic term counteracts the unboundedness of the potential. 
As we will see, the information about the boundary conditions also  appears in the classical Hamiltonian constraint 
(\ref{eq:hamiltonianconstfull}) of $2$d projectable HL gravity with a wormhole interaction.   

In the next section, we will show that the path-integral (\ref{eq:pathintegralfullg}) can be obtained quantizing the $2$d projectable HL gravity with a wormhole interaction.

\section{Wormhole interaction in $2$d projectable HL gravity}\label{sec:wormholeinhl}

We consider  $2$d projectable HL gravity with a space-like wormhole interaction given by the action: 
\begin{align}
I_w 
&= 
\frac{1}{\kappa}
 \int ds dx 
 N(s)\sqrt{h(s,x)}
\left(
(1-\eta)K^2(s,x) -2 \tilde\lambda 
\right)\notag \\ 
 &\ \ \ + \beta 
\int ds N(s) \int dx_1 dx_2 \sqrt{h(s,x_1)} \sqrt{h(s,x_2)}\ , 
\label{eq:shlw}
\end{align}
where $\beta$ is a dimensionfull coupling constant. 
One can show that the action (\ref{eq:shlw}) is invariant under the FPD (\ref{eq:fpd}) with the projectable lapse function, $N=N(t)$. 
The bi-local interaction in eq.\ (\ref{eq:shlw}) relates two distinct points at an equal time.  
The action is a simplified version of the general bi-local action suggested in the article\cite{hawking}, made possible because HL 
gravity is invariant only under the FPD (\ref{eq:fpd}) and not the full set of diffeomorphisms.

Following the same procedure explained in the section \ref{sec:2dhl}, we quantize the system with the action (\ref{eq:shlw}). 
Introducing a conjugate momentum of $\sqrt{h}$ as $\pi$ as before, 
we move on to the canonical formalism, 
and the Hamiltonian is 
\begin{equation}
H = \int dx\ \left[
N_1 \mathcal{C}^1 (s,x)
+N \mathcal{C} (s,x)
\right]\ ,
\label{eq:extendedclassicalh}
\end{equation}
where 
\begin{align}
\mathcal{C}^1 (s,x) 
&= - \frac{\partial_1 \pi (s,x)}{\sqrt{h (s,x)}}\ , 
\label{eq:momentumconstfull} \\
\mathcal{C} (s,x) 
&=  
\frac{\kappa}{4(1-\eta)} \pi^2 (s,x) \sqrt{h(s,x)} 
+ \frac{2}{\kappa}\tilde \lambda \sqrt{h (s,x)} 
- \beta \sqrt{h (s,x)} \int dx_2 \sqrt{h (s,x_2)}\ . 
\label{eq:hamiltonianconstfull}
\end{align}

If we solve the momentum constraint (\ref{eq:momentumconstfull}) at the classical level, 
the system again reduces to the $1$d system with the Hamiltonian: 
\begin{equation}
H= N(s) \left( \frac{\kappa}{4(1-\eta)} \pi^2(s) \ell (s) + \frac{2}{\kappa} \tilde \lambda \ell (s) - \beta \ell^2 (s)  \right)\ ,
\label{eq:1dhamiltonianfull}
\end{equation}
where $\ell (s) := \int dx\ \sqrt{h(s,x)}$. 
Hereafter choosing the correct sign for the kinetic term, i.e. $\eta<1$, 
we use the parametrization (\ref{eq:lambdacdt}) with positive $\lambda$ as before.   

The classical Hamiltonian constraint (\ref{eq:hamiltonianconstfull}) can be solved as 
\begin{equation}
\pi^2 = - \lambda +  \beta  \ell \ge 0 \ ,
\label{eq:c2}
\end{equation}
for $\sqrt{\lambda}\ell \ge 1/\xi$ with $\xi:=\beta/\lambda^{3/2}$, 
and otherwise, the classical Hamiltonian constraint (\ref{eq:hamiltonianconstfull}) requires $\ell =0$ on the constraint surface. 
As in the $\beta=0$ case, when quantizing the system based on the path-integral, we don't have any problem with respect to the quantization around $\ell (s)=0$.   

Simply repeating the procedure in the section \ref{sec:2dhl}, one can show that the amplitude (\ref{eq:pathintegralfullg}) can be precisely recovered
by quantizing the $2$d projectable HL gravity with the bi-local interaction based on the path-integral, 
if $\beta = g_s$.

\section{Discussions}\label{sec:discussions}

We have canonically quantized the $2$d projectable HL gravity with a simple space-like wormhole interaction, 
and shown that the quantum Hamiltonian is equivalent to the one-body quantum Hamiltonian that includes contributions coming from all wormholes and baby universes obtained in the string field theory for CDT, 
if $g_s = \beta $ and $\lambda = \tilde\lambda/(2(1-\eta)) $ where $\beta>0$, $\lambda > 0$ and $\eta <1$.

Let us consider the classical Hamiltonian constraint $\mathcal{C}$ (\ref{eq:hamiltonianconstfull}) in the parameter region above.  
When $\sqrt{\lambda}\ell \ge 1/\xi$ where $\xi$ is a dimensionless quantity defined by $\xi := g_s/ \lambda^{3/2}$, 
the constraint surface is given by     
$
\pi^2 = - \lambda+ g_s \ell \ge 0
$. 
However, when $\sqrt{\lambda}\ell < 1/\xi$, the only allowed solution is $\ell=0$. 
In the case of the $2$d projectable HL gravity ($\beta = g_s = 0$), with the parameter region corresponding to $2$d CDT, i.e. $\lambda>0$, 
the only solution to the classical Hamiltonian constraint is $\ell=0$. 
Therefore, when $\sqrt{\lambda}\ell < 1/\xi$, the classical solution of the $2$d projectable HL 
with the wormhole interaction will be close to that of the $2$d projectable HL gravity, 
if one sits in the parameter region above; they can be quite different when $\sqrt{\lambda}\ell \ge 1/\xi$. 
Such a relation also holds at the quantum level. As shown in the article\cite{Ambjorn:2021wdm}, when $\sqrt{\lambda}\ell < 1/\xi$
the eigenfunctions of the one-body quantum Hamiltonian including all wormholes and baby universes, $H_{-1}$ (\ref{eq:onebodyfullh}), 
can be well approximated by the eigenfunctions of the quantum Hamiltonian of $2$d CDT without wormholes and baby universes, $H^{(0)}_{-1}$ (\ref{eq:hamiltonians}).
On the other hand, when $\sqrt{\lambda}\ell \ge 1/\xi$, their behaviors are quite different, 
and in this case, in order for the theory to be well-defined, the unbounded nature of the potential in $H_{-1}$ will be 
counteracted by the kinetic term. This  balance between the kinetic and potential terms is precisely what is reflected in
the classical Hamiltonian constraint (\ref{eq:hamiltonianconstfull}). 
This is also consistent with the boundary conditions on $\ell$ at infinity in the path-integral (\ref{eq:measure3}).

The picture of creation and annihilation of baby universes and wormholes   is conceptually straightforward in the string field
theory for CDT \cite{Ambjorn:2008ta}. Nevertheless  it is somewhat surprising that one from this can derive an effective 
one-body Hamiltonian which can describe  propagation of a single spatial universe, i.e.\ the propagation where the spatial universe starts
with the topology of a circle and at a later time $t$ has the same topology, but where it in the intermediate times  is allowed
to split in two and either one part disappears in the vacuum (a baby universe), or the two parts join again at a later time (then 
changing the spacetime topology). This process of joining and splitting  can be iterated at  intermediate times 
and using string field theory we can perform the summation of all iterations and  derive the effective one-body 
Hamiltonian (\ref{eq:onebodyfullh}). From the point of view of unitary evolution (or using Euclidean time, semigroup evolution, to 
be more precise),  as given in eq.\ (\ref{eq:cylinder3}), it is difficult to understand how a complete set of intermediate states can 
both be given by the one spatial universe states $| \ell \rangle$ and by the complete multi-universe Fock states of the string field theory (for a recent discussion of this issue see i.e.\ \cite{don}). 
However, this seems to be the case by explicit calculation, and we find it even more surprising that the simplest wormhole interaction term added to the classical action as in eq.\ (\ref{eq:shlw}) leads to precisely the same single universe quantum Hamiltonian as found in the string field theory for CDT.


\end{document}